\documentclass{elsart3}

 \usepackage{epsfig}

\usepackage{amssymb}

\def\nle{\ \raise.3ex\hbox{$<$}\kern-0.8em\lower.7ex\hbox{$\sim$}\ }
\def\nge{\ \raise.3ex\hbox{$>$}\kern-0.8em\lower.7ex\hbox{$\sim$}\ }
\def\chitwo{\chi''(\omega; t)}
\def\lcrh{L_h}
\def\lovT{L_{\Delta T}}
\def\Rcreff{R_{\rm cr}^{\rm eff}}
\def\Tc{T_{\rm c}}
\def\tcr{t_{\rm cr}}
\def\tcrdn{t_{\rm cr}^{\rm n}}
\def\tcrup{t_{\rm cr}^{\rm p}}
\def\Tf{T_{\rm f}}
\def\tma{t_{\rm max}}
\def\tmi{t_{\rm min}}
\def\tw{t_{\rm w}}

\def\twdn{t_{\rm w}^{\rm n}}
\def\twup{t_{\rm w}^{\rm p}}
\def\fcm{M_{\rm FC}}
\def\trm{M_{\rm TR}}
\def\zfcm{M_{\rm ZFC}}
\def\FeMnTiO{Fe$_{0.5}$Mn$_{0.5}$TiO$_3$}

\begin{document}

\begin{frontmatter}



\title{Aging phenomena in spin glasses: theory, experiment, and simulation}


\author{H. Takayama}

\address{Institute for Solid State Physics, University of Tokyo,
 Kashiwa-no-Ha 5-1-5, Kashiwa 277-8581, Japan}

\begin{abstract}
We study numerically temperature-shift and field-shift aging
 protocols on the 3-dimensional (3D) Ising Edwards-Anderson (EA) 
spin-glass (SG) model focusing on respectively the
 temperature-chaos nature and the stability under a static field of the
 SG phase. The results of the latter strongly support
 the droplet theory which predicts the instability of the SG phase
 under the field. They are also discussed in relation with the
 experimental studies.    
\end{abstract}
\end{frontmatter}

\section{Introduction}
\label{}

More than a decade the dispute on the nature of ordering in spin glasses
between the mean-field theory based on the replica symmetry breaking in
Parisi's solution of the infinite-ranged model and the droplet theory
based on a phenomenological scaling argument on the short-ranged model
has remained unsettled \cite{review1}. Currently, the following two explicit
predictions by the droplet theory \cite{FH-88-EQ,BM-chaos} 
have been extensively studied. 1) The equilibrium SG 
states at two temperatures with difference 
$\Delta T$ are uncorrelated in length scales larger than the overlap
length $\lovT$, i.e., the so-called temperature($T$)-chaos
nature of the SG phase, and 2) in the equilibrium and
thermodynamic limits the SG phase is broken by a static field
$h$ of infinitesimal strength, thereby introduced is the crossover
length $\lcrh$. The latter separates characteristics of droplet
excitations by their size $L$, such that they are dominated by the
Zeeman energy for $L>\lcrh$ and by the SG free energy gap
for $L<\lcrh$.
 
One may imagine that the existence of such a characteristic length 
$\lovT\ (\lcrh)$ could be easily checked by the 
temperature($T$)-shift (field($h$)-shift) aging protocol, where the 
temperature 
(field) is changed once on a way to equilibrium. However, this is rather
hard because of the extremely slow spin dynamics in the SG state:  
within a time-window of relative magnitude of $10^{4 \sim 5}$, which is
the case not only in simulations but also in ordinary experiments, the
SG correlation length, or mean domain size, $R(t)$, is
considered to grow only by a factor two or even less as will be discussed
below. Therefore, no simulation nor experiment on such an aging protocol
alone has not yet succeeded to settle the problem mentioned above. 
A possible strategy to overcome this difficulty is to examine a set of 
$T$-\ ($h$-) shift protocols with $\lovT\ (\lcrh)$ systematically changed 
relatively to $R(t)$ within the available time-window, to look for a 
possible scaling behavior in the length scales involved, and to check 
whether the behavior is compatible with the droplet theory or 
not \cite{PYN-02}.

In the present work we numerically study aging dynamics in the 3D Ising
EA SG model with Gaussian nearest-neighbor interactions
with zero mean and variance $J$. The latter is used as the unit of
energy and temperature. In this unit the SG transition
temperature is 
estimated as $\Tc \simeq 0.95$ \cite{Marinari-cd98-PS,MariCamp}. 
Along the above-mentioned strategy
we have examined a number of $T$- and $h$-shift processes with various 
$\Delta T$ and $h$ as well as the (waiting) time, $\tw$, of the
shift. For further details of the model and the Monte Carlo (MC) method
of simulation the reader may refer to our previous 
works \cite{ours1,ours2,oursSupp,oursLoren,ours3}. 

In the next section we investigate the $T$-chaos nature through
$T$-shift processes, and in \S 3 the stability of the
SG phase under a field through $h$-shift processes. The
last section is devoted to discussions.


\section{$T$-shift aging processes}

\begin{figure}
\begin{center}
 \resizebox{0.45\textwidth}{!}{\includegraphics{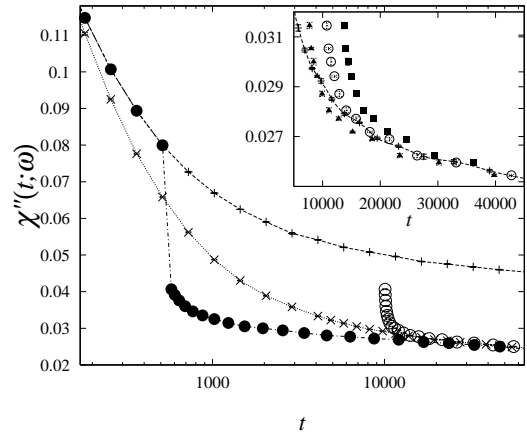}}
\caption{$\chitwo$ in a negative $T$-shift process with $\twdn=512$,
 $T_1=0.7$,  $T_2=0.4$, and $t_{\omega}=2\pi/\omega=64$ (solid
 circles). The two continuous curves represent $\chitwo$ in the
 isothermal aging at $T=T_1$ and $T_2$. The open circles are the
 properly shifted branch of $\chitwo$ at $t>\twdn$. Shown in the inset
 are how the shifted branches merge to the $T_2$-isothermal curve (see
 the text).}
\label{fig:T-shift}
\end{center}
\end{figure}

\begin{figure}
\begin{center}
 \resizebox{0.45\textwidth}{!}{\includegraphics{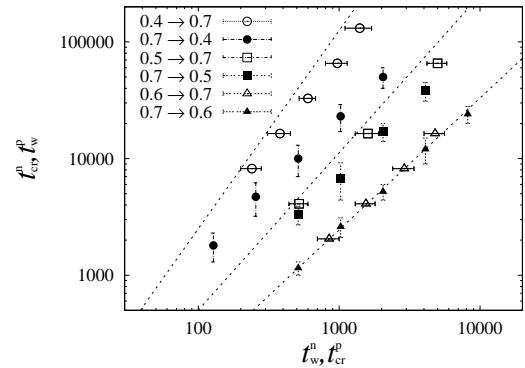}}
\caption{$\tcrdn$ ($\twup$) vs. $\twdn$ ($\tcrup$) for the
 negative (positive) $T$-shift processes between three pairs of
 temperatures. The lines  represent the expected behavior from
 Eq.(\ref{eq:tcross}) combined with  Eq.(\ref{eq:RT(t)}) as explained in
 the text.
}
\label{fig:tcrossT}
\end{center}
\end{figure}

Following the idea of `twin-experiments' in \cite{PYN-02}, we examine
here negative (positive) $T$-shift aging processes. A system is
instantaneously quenched at time $t=0$ from 
$T=\infty$ to $T_1\ (T_2)$ below $\Tc$, the temperature is decreased
(increased) to $T_2 \ (T_1)$ at $t=\twdn \ (\twup)$, and then let the
system to age at $T_2\ (T_1)$. Throughout
the process we measure the imaginary part of the ac susceptibility of
frequency $\omega$, $\chitwo$ \cite{ours2}. 
Its typical behavior in a negative
$T$-shift process is shown in Fig.~\ref{fig:T-shift}. By the $T$-shift
at $t=\twdn$, $\chitwo$ rapidly decreases, and it merges to the
$T_2$-isothermal curve from below. If, however, the branch of $\chitwo$
at $t>\twdn$ is shifted to the {\it right} by a proper amount, 
$t_{\rm sh}$, as indicated by the open circles in the figure, it merges
to the $T_2$-isothermal curve from above at $t=t_{\rm mr}$.
The inset of the figure explains how to determine $t_{\rm sh}$: when an
amount of the shift is smaller than $t_{\rm sh}$, the shifted branch
crosses with the $T_2$-isothermal curve (triangles in the inset), while
a larger shift delays the merging significantly (squares). Here we call 
the time required for $\chitwo$ to merge to the isothermal curve in this
sense the {\it crossover time} and denote it as $\tcrdn$, i.e., 
$\tcrdn = t_{\rm mr}-(\twdn+t_{\rm sh})$. [This characteristic time is 
called the effective age in~\cite{PYN-02} and the effective waiting time 
in~\cite{ours3}.] For a positive $T$-shift process we can similarly 
define the crossover time $\tcrup$ \cite{ours3}. A set of results for 
$\tw$ and $\tcr$ thus obtained are shown in Fig.~\ref{fig:tcrossT}.

After the first quench of a negative $T$-shift process, the SG ordering, 
or the mean domain size $R(t)$, grows up to $R(\tw)$ at $t=\tw$. By the 
$T$-shift $R(t)$ does not exhibit a discontinuous change, but continues 
to grow now with the growth rate at $T_2$ \cite{oursSupp}. However, it 
takes time $\tcr$ for $\chitwo$ to merge to its $T_2$-isothermal curve. 
We suppose that in this time range, which we call the {\it transient 
regime}, subdomains of a mean size $L(t-\tw)$ in local equilibrium of 
$T_2$ grows within each domains of a mean size $R(\tw)$ in local 
equilibrium of $T_1$. If $R(\tw)$ is enough smaller than the overlap 
length $\lovT$ introduced in \S 1, we naively expect that $L(t-\tw)$ 
catches up $R(\tw)$ at $t-\tw \simeq \tcr$, or, 
\begin{equation}
        L(\tcr) = R_{T_2}(\tcr) = R_{T_1}(\tw),
\label{eq:tcross}
\end{equation}
to hold, where $R_T(t)$ is the growth law of SG domains in the
isothermal aging at $T$. By the first equality above we imply that the
growth mechanisms of $L(t-\tw)$ at $T$ and $R_T(t)$ are common, 
i.e., they are
due to thermal activation processes. The same equation as above but 
with $T_1$ and $T_2$ interchanged is expected to hold for the 
corresponding positive $T$-shift process. This expectation is in fact 
confirmed by the result in Fig.~\ref{fig:tcrossT} that the data $\tcrdn$ 
vs. $\twdn$ and $\twup$ vs. $\tcrup$ lie on a common curve for a small 
$\Delta T = T_1-T _2\ (=0.1)$. We call the interpretation so far 
described the {\it cumulative memory scenario} for $T$-shift processes 
with small $|\Delta T|$. 

For later discussions let us here mention explicit expressions for the 
growth law of $R_T(t)$. Within the accuracy of simulations, the numerical
results of $R_T(t)$ \cite{ours1,Kisker-96,Marinari-growthLaw} are well 
fitted either to the logarithmic law proposed in the droplet 
theory \cite{FH-88-NE}    
\begin{equation}
   R_T(t)/l_0  \sim [{T \over \Delta_{\rm g}}{\rm ln}(t/t_0)]^{1/\psi}, 
\label{eq:Rt-FH}
\end{equation}
or to a power law
\begin{equation}
        R_T(t)/l_0 \simeq (t/t_0)^{1/z(T)}.
\label{eq:RT(t)}
\end{equation}
Above, $l_0$ and $t_0$ are the characteristic length and time scales, and 
$\Delta_{\rm g}$ and $\psi$ the characteristic scale of energy barrier of 
droplet excitations and the associated exponent, respectively. In 
Eq.(\ref{eq:RT(t)}) we set $l_0=$ 1 lattice distance and $t_0=$ 1 MC step 
per spin, and the exponent $1/z(T)$ is given by \cite{ours1}
\begin{equation}
1/z(T) \simeq bT/\Tc,
\label{eq:z_T}
\end{equation}
with $b \simeq 0.16$ in the range $0.7 \nge T/\Tc \nge 0.4$. 

For a large $\Delta T \ (=0.3)$, the sets of data of negative and
positive $T$-shift processes in Fig.~\ref{fig:tcrossT} differ
significantly from each other. Also the second equality of
Eq.(\ref{eq:tcross}) is violated for the negative $T$-shift process shown 
in Fig.~{\ref{fig:T-shift}}. We tentatively attribute this deviation
from the cumulative memory scenario to the temperature-chaos nature of
the equilibrium SG phase described in \S 1. If $\lovT$ is smaller than 
$R_{T_1}(\tw)$, then in the time range after the $T$-shift in which 
$L(t-\tw) > \lovT$ is satisfied, aging dynamics looks as if the system is 
already in the $T_2$-isothermal aging state. This is because, by 
definition of $\lovT$, the longer range order than $\lovT$ developed at 
$T=T_1$ is irrelevant to the SG order at $T=T_2$. Therefore $\tcr$ is 
given by the condition $L(\tcr)\simeq \lovT$, and is independent of $\tw$ 
in this case.  

However, $\tw$ in the present simulation is not large enough to
reproduce such a drastic phenomenon. We consider that the negative
$T$-shift process observed here is in the {\it weak chaos regime}
proposed by J\"onsson {\it et al} \cite{PYN-02-R}. We also note that the
second equality of Eq.(\ref{eq:tcross}) is little violated in the
positive $T$-shift with $\Delta T =0.3$. This asymmetric appearance of
the chaos effect contradicts the prediction of the droplet 
theory \cite{FH-88-EQ} as well as the recent experiment on the
Heisenberg spin glass AgMn \cite{PYN-02}.

\section{$h$-shift aging processes}

The $h$-shift processes we mainly discuss here are as follows. After 
instantaneous quench to $T$, the system is aged under a zero field until 
$t=\tw$, when a field $h$ is switched on. We then observe the induced 
(zero-field-cooled) magnetization  $M\ (=\zfcm)$ and $\chitwo$ in the 
whole process. We study them exclusively at $T=0.6$ where the aging 
dynamics is considered not affected by the critical dynamics close to 
$\Tc$. 

\begin{figure}
\begin{center}
 \resizebox{0.45\textwidth}{!}{\includegraphics{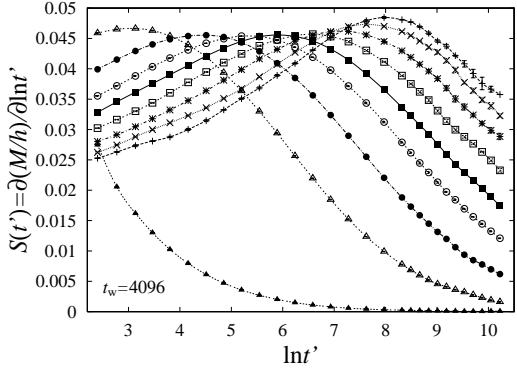}}
\end{center}
\caption{ $S(t')$ with $\tw=4096$ plotted vs ln$t'$. The data are for
 $h=$0.1, 0.2, 0.3, 0.4, 0.5, 0.6, 0.75, 1.0 and 2.0 from right to
 left.
 }
\label{fig:Soft-up}
\end{figure}

\begin{figure}
\begin{center}
\resizebox{0.4\textwidth}{!}{\includegraphics{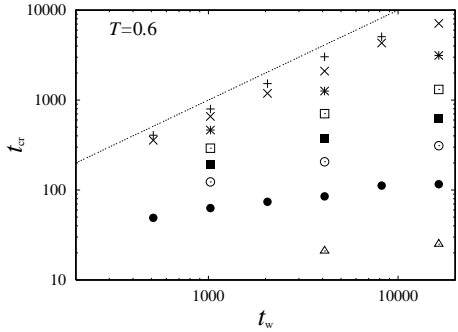}}
\end{center}
\caption{ Plot $\tcr$ vs. $\tw$. The data points
 are for $h=$0.1, 0.2, 0.3, 0.4, 0.5, 0.6, 0.75 and 1.0 from top to
 bottom, and the line represents $\tcr=\tw$. }
\label{fig:tcr-tw}
\end{figure}

It is well established that, at least for a small $h$, the logarithmic
growth rate of $M$ defined by 
$S(t') = \partial (M/h) /\partial {\rm ln}t'$ with $t'=t-\tw$
exhibits a peak at $t'\sim \tw$. More generally, the peak position of 
$S(t')$ plays the same role as $\tcr$ discussed in the previous section. 
It gives us a characteristic time scale of crossover from the isothermal 
aging state under $h=0$ to that under a finite $h$, and so is denoted 
also by $\tcr$.  In Fig.~\ref{fig:Soft-up} we show $S(t')$ with 
$\tw=4096$ for various values of $h$. For a small $h\ (=0.1)$, as 
mentioned just above, $\tcr \simeq \tw$ is ascertained. When $h$ 
increases, however, $\tcr$ significantly decreases. In 
Fig.~\ref{fig:tcr-tw} the resultant $\tcr$ for various $h$ and $\tw$ are 
presented. 

Now we convert the relations between $\tcr$ and $\tw$ into those between 
the characteristic length scales $\Rcreff = R_T(\tcr,h)$ and $R_T(\tw)$. 
The effective crossover length $\Rcreff$ is the mean size of subdomains 
grown under $h$ in the period of $\tcr$ and has to be estimated by the 
growth law under $h$. Since, however, it is quite time-consuming to 
simulate the latter due to the presence of the paramagnetic component, we 
approximate it by $R_T(\tcr,h=0)$, i.e., by Eq.(\ref{eq:RT(t)}) in the 
present work. 

\begin{figure}
\begin{center}
\resizebox{0.45\textwidth}{!}{\includegraphics{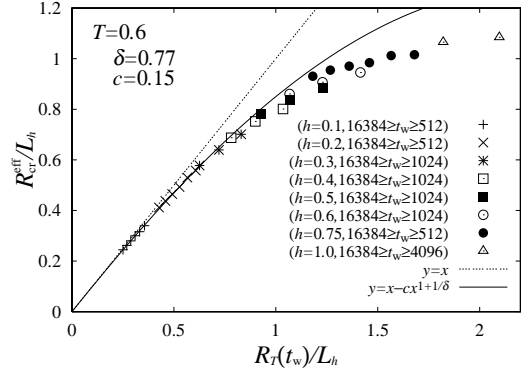}}
\end{center}
\caption{$\Rcreff/\lcrh$ vs $R_T(\tw)/\lcrh$. The data at small $x$ are 
fitted to $y=x-cx^{1+1/\delta}$ with $c=0.15$, which is the formula 
proposed in the weak chaos scenario in \cite{PYN-02-R}. }
\label{fig:Rcreff-scal}
\end{figure}

The obtained sets of $\Rcreff$ and $R_T(\tw)$ for different $h$ are 
normalized by the crossover length $\lcrh$ introduced in \S 1. It is 
given by \cite{FH-88-EQ}       
\begin{equation}
\lcrh/l_0 \simeq  (h/\Upsilon)^{-\delta}, 
\label{eq:lcrh}
\end{equation}
with $\delta = ({d \over 2} - \theta)^{-1}$, where $\Upsilon$ is the
unit of free-energy gap. In evaluating explicitly $\lcrh$ by 
Eq.(\ref{eq:lcrh}) we simply put $\Upsilon$ unity and $\delta=0.77$ 
with $d=3$ and $\theta=0.2$ \cite{ours1,BM-theta}. 
The resultant scaling plot of $\Rcreff/\lcrh$ vs. $R_T(\tw)/\lcrh$ is
presented in Fig.~\ref{fig:Rcreff-scal}. One sees in the figure that most 
of the data tend to lie on a single curve. If we take into account the 
fact that we have not adjusted the parameters involved at all, the 
scaling behavior obtained is quite satisfactory.  

We interpret the scaling behavior shown in Fig.~\ref{fig:Rcreff-scal} as
dynamical crossover from the SG aging state to the paramagnetic state. 
Actually, in the range $x \ll 1$, where $R_T(t)$ measured within the time 
window of our simulation is much smaller than $\lcrh$, various properties 
which are usually regarded as those of the SG state are observed. Among 
them are the so-called $\omega t$-scaling of the ac susceptibility in the 
isothermal aging \cite{review1} 
and the sum rule $\zfcm(t) + \trm(t) = \fcm(t)$ 
\cite{sum-rule}, where $\fcm$ is the induced magnetization observed when 
$h$ is applied just after the quench to $T$ and $\trm$ is the 
thermoremanent magnetization observed after $h$ in the field-cooled
process is switched off at $t=\tw$. In the range $x \nge 1$ corresponding 
to a large field of $h \nge 1.0$, on the contrary, the complete 
saturation of $\zfcm$ and $\fcm$ to a common value is observed within the 
time window of our simulation. This is the expected property of the 
paramagnetic state. Combined with these observations which will be 
discussed elsewhere, we conclude that the result of 
Fig.~\ref{fig:Rcreff-scal} is a strong support of the droplet picture
for the $h$-shift aging process and so the instability of SG
state in the equilibrium and thermodynamic limits.

Lastly let us try to extend the scaling behavior obtained above by our 
simulation on the Ising EA model to the one in real Ising spin glasses 
such as \FeMnTiO, focusing on the crossover point defined by the condition 
$x=R_T(\tw)/\lcrh=1$. Here we use the power growth law of 
Eq.(\ref{eq:RT(t)}) with $t_0=10^{-12}$s. Then, with the same 
approximations, such as $R_T(t,h) \simeq R_T(T,h=0)$, as those already 
introduced, the condition is reduced to 
\begin{equation}
h^{-\delta} \simeq (t/t_0)^{-bT/\Tc}.
\label{eq:length-h}
\end{equation}
Also using $b=0.16$ and $\delta=0.77$ as in 
the simulation, we obtain, for example,  $h \simeq 0.6$T for $T/\Tc=0.6$ 
and $t=10^2$~s. A set of these figures is compared with $h \simeq 1.3$~T, 
$T_{\rm AT}/\Tc \simeq 0.6$ and $t_{\rm me} \sim 10^2$~s, which are 
picked up from the paper by Katori and Ito \cite{KatoriIto94}, who 
determined the critical temperature $T_{\rm AT}$ under $h$ of \FeMnTiO\ 
as the one where the $\zfcm$ starts to deviate from $\fcm$. In a certain 
range of $T_{\rm AT}$, they found the relation 
$h \propto (1-T_{\rm AT}/\Tc)^\alpha$ with $\alpha \simeq 1.5$ which is 
the expected behavior of the de Almaid-Thouless phase transition 
\cite{AT-line} in the infinite-ranged SG model. However, the rather good 
coincidence of the above two sets of vaules, combined with the dependence 
of $h$ on $T$ from Eq.(\ref{eq:length-h}) with a fixed $t$ equal to the 
measuring time $t_{\rm me}$ in their temperature sweeping 
experiment (not shown), suggests that $T_{\rm AT}$ they measured 
could be interpreted as the dynamical crossover point with 
$t_{\rm me} \simeq \tcr$ . 

Another experiment on \FeMnTiO\ under large static field $h$ was done by
Mattsson {\it et al.} \cite{Mattsson95}, who measured the ac
suceptibility. They found two regimes in its behavior which are
separated by the the freezing temperature $\Tf(\omega,h)$. Although what 
they measued is droplet excitations almost in equilibrium but not 
domains in aging, we may use Eq.(\ref{eq:length-h}) with $t=t_\omega = 
2\pi/\omega$, i.e., $\Tf$ is the temprature at which the mean size of
droplets which can respond to the ac field of period $t_\omega$ is equal 
to $\lcrh$. Then Eq.(\ref{eq:length-h}) yields, for 
example, $h \simeq 3.0$~T with $t_\omega =10^{-2}$~s and $\Tf/\Tc=0.5$,
while the corresponding experimental value is $h=2.2$~T. 

Although we do not want to insist strongly the semi-quantitative
agreement of the experimental results with a simple extension of the 
simulational results, we may claim from the above 
argument that $\tcr$ for field-shift processes under $h$ of a few Oe is 
beyond the astronomical time.

\section{Discussions}

On the interpretation of aging phenomena in spin glasses by means of the
droplet picture, the length scales of certain domains and droplets play
a central role. In the above arguments, they are $\lovT$, $\lcrh$, 
$R_T(\tw)$,  and $L(t-\tw)$. In the numerical study so far carried out 
\cite{ours1,Kisker-96,Marinari-growthLaw}, $R_T(t)$ is replaced by the 
coherence length of the real replica overlap function which has been 
explicitly evaluated. Since the overlap function is obtained by averaging 
over sites in each sample and then over samples, detailed properties of 
each domains, such as whether they are fractal or compact, are averaged out, 
and so they are beyond the scope of the present simulation. Neither the 
nature of subdomains has been directly pursued by simulation yet. But its 
introduction with the {\it subdomains-within-domain scenario} 
\cite{oursSupp,ours3} helps us very much to interpret the observed 
phenomena in the transient regime after the $T$- or $h$-shift as well as 
the {\it memory effect} observed in the $T$- or $h$-cycling processes 
\cite{review1}. For example, even in the cummulative regime of 
Fig.~\ref{fig:Rcreff-scal} with $\Rcreff \simeq R_T(\tw)$, 
rejuvenation-like behavior (or a jump-up) in $\chitwo$ has been observed 
both in experiment \cite{h-Eric95} and the present simulation 
(not shown). 

Another fundamental ingredient of the analysis based on the droplet
picture is the growth law of $R_T(t)$, which is frequently used as the 
conversion formula from the time scale to the length scale. However it is 
hard to accurately determine it due to the following circumstances. Let us 
consider the ratio $r = R_T(\tma)/R_T(\tmi)$, where $\tma \ (\tmi)$ is 
the maximum (minimum) time scale of observation. The results of our 
simulation with $\tmi \simeq 10$ and $\tma \simeq 10^5$ yield, say at 
$T/\Tc=0.6$, $r \simeq 2.3$. If Eqs.(\ref{eq:RT(t)}) and (\ref{eq:z_T}) 
with $t_0=10^{-12}$~s and $b=0.16$ are extended to the time scale of 
ordinary experiments, i.e., $\tmi \sim 10$~s and $\tma \sim 10^5$~s, the 
identical $r$ is obtained. When the logarithmic law of 
Eq.(\ref{eq:Rt-FH}) is applied to the experimental data, $r$ becomes even 
smaller \cite{PYN-02}. We may say that, both in experiments and 
simulations, we measure a very limited portion of an equilibration 
process, at least the length scale of SG ordering is concerned. This is 
the reason of difficulty in determining the growth law of $R_T(t)$ 
accurately. 

In order to investigate the stability of the equilibrium SG state against 
perturbations such as the temperature change $\Delta T$ and 
a static field $h$, thereby overcoming the above-mentioned 
difficulty, we have carried out a set of shift-aging protocols where the 
perturbation of various strengths, $P\ (\Delta T$ or $h)$, is applied 
after the waiting time $\tw$ which are also systematically changed. For 
each process specified by $(P,\tw)$, we have extracted the crossover 
time $\tcr$ between the state aging to an unperturved SG state and the 
one to the perturbed SG, or non-SG state. The length scales 
$\Rcreff(P;\tcr)$ associated with $\tcr$ are then found to be scaled 
nicely by the characteristic length scale associated with $P$. 
This strategy, first carried out in ~\cite{PYN-02}, is 
expected to be most efficient to judge the stability of the equilibrium 
SG state by actual experiments.

Irrespectively of the strategy mentioned above, experiments to extract 
$R_T(t)$ with referring to another length scale independently of the 
growth law itself are certainly very much
required. To our knowledge, however, there has been only two such
experiments which use a length scale associated with the static field
effect. One is due to Mattsson {\it et al} \cite{Mattsson95} already
mentioned above and the other due to Joh {\it et al} \cite{Joh-Or}.
The latter result prefers the power law of Eq.(\ref{eq:RT(t)}) with the
value of $b$ in $1/z(T)$ nearly equal to our numerical results. 
From this point of view,
an experiment on SG fine particle systems may be of interest.

To conclude we have presented and discussed the results of our simulation 
on the $T$- and $h$-shift aging processes in the 3D Ising EA SG model. In the 
$T$-shift process only a precursor of the temperature-chaos effect has been
observed, while the results on the $h$-shift process strongly support 
the droplet picture that the SG state under a finite field is 
unstable in the equilibrium and thermodynamic limits, though it takes 
astronomic time for the SG state to be broken if $h$ is small.

I would like to sincerely thank K. Hukushima, P. J\"onsson, H. Yoshino, 
and E. Vincent 
for their fruitful discussions. The present simulation has been 
performed by using the computor facilities at Supercomputer Center,
Institute of Solid State Physics, the University of Tokyo.


\begin{thebibliography}{00}

\bibitem{review1}
See the papers in {\it Spin glasses and random fields}, ed., A.P. Young,
(World Scientific, Singapore, 1997). 

\bibitem{FH-88-EQ}
D.S. Fisher and D.A. Huse, Phys. Rev. B  {\bf 38} (1988) 386.

\bibitem{BM-chaos}
A.J. Bray and M.A. Moore,
Phys. Rev. Lett. {\bf 58} (1987) 57.

\bibitem{PYN-02}
P.E. J\"onsson et al.,
Pyhs. Rev. Lett. {\bf 89} (2002) 097201.

\bibitem{Marinari-cd98-PS}
 E. Marinari et al., 
Phys. Rev. B {\bf 58} (1998) 14852.

\bibitem{MariCamp}
 P.O. Mari and I.A. Campbell,
 Phys. Rev. E {\bf 59} (1999) 2653.
 
\bibitem{ours1}
T. Komori et al., 
J. Phys. Soc. Jpn. {\bf 68} (1999) 3387.

\bibitem{ours2}
T. Komori et al., 
J. Phys. Soc. Jpn. {\bf 69} (2000) 1192.

\bibitem{oursSupp}
T. Komori et al., 
J. Phys. Soc. Jpn. {\bf 69} Suppl. A (2000) 355.

\bibitem{oursLoren} 
L.W. Bernardi et al., 
Phys. Rev. Lett. {\bf 86} (2001) 720.

\bibitem{ours3}
H. Takayama and K. Hukushima,
J. Phys. Soc. Jpn. {\bf 71} (2002) 3003.

\bibitem{Kisker-96}
 J. Kisker et al., 
Phys. Rev. B  {\bf 53} (1996) 6418.

\bibitem{Marinari-growthLaw}
 E. Marinari et al., 
J. Phys. A  {\bf 31} (1998) 2611.

\bibitem{FH-88-NE}
D.S. Fisher and D.A. Huse, Phys. Rev. B  {\bf 38} (1988) 373.

\bibitem{PYN-02-R}
P.E. J\"onsson et al., 
Pyhs. Rev. Lett. {\bf 90} (2003) 059702.

\bibitem{BM-theta}
 A.J. Bray and M.A. Moore,
J. Phys. C  {\bf 17} (1984) L463.

\bibitem{sum-rule}
P. Nordblad et al., 
L. Magn. Magn. Mat. {\bf 54} (1986) 185.

\bibitem{KatoriIto94}
H. Aruga Katori and A. Ito, 
J. Phys. Soc. Jpn. {\bf 63} (1994) 3122.

\bibitem{Mattsson95}
J. Mattsson et al., 
Phys. Rev. Lett. {\bf 74} (1995) 4305.

\bibitem{AT-line}
J.R.L.de Almaida and D.J. Thouless,
J. Phys. A {\bf 11} (1978) 983.

\bibitem{h-Eric95}
E. Vincent et al., 
Phys. Rev. B {\bf 52} (1995) 1050.

\bibitem{Joh-Or}
Y. G. Joh et al., 
Phys. Rev. Lett. {\bf 82} (1999) 438, and J. Phys. Soc. Jpn. {\bf 69} 
Suppl. A (2000) 215.


\end{thebibliography}
\end{document}